\documentclass[11pt]{article}

\usepackage{a4}
\usepackage{latexsym,amssymb}
\usepackage{citesort}

\usepackage{url}
\usepackage{ntheorem}

\newtheorem{Theorem} {\sc  Theorem\rm} [section]

\newtheorem{Proposition} [Theorem] {\sc  Proposition\rm}

\newcommand{\zphione}{\chi}

{\catcode `\@=11 \global\let\AddToReset=\@addtoreset}
\AddToReset{equation}{section}

\newcommand{\zh} {\mathring{h}} 
\newcommand{\zX}{{\mathring{Z}}}

\newcommand{\zR}{\mathring{R}}

\newcommand{\zmcH}{{\cal N}}

\DeclareFontFamily{OT1}{rsfs}{}
\DeclareFontShape{OT1}{rsfs}{m}{n}{ <-7> rsfs5 <7-10> rsfs7 <10-> rsfs10}{}
\DeclareMathAlphabet{\mycal}{OT1}{rsfs}{m}{n}
\def\scri{{\mycal I}}%
\newcommand{\zA}{\mathring{A}}%

\newcommand{\bel}[1]{\begin{equation}\label{#1}}
\newcommand{\bea}{\begin{eqnarray}}
\newcommand{\bean}{\begin{eqnarray}\nonumber}
\newcommand{\beal}[1]{\begin{eqnarray}\label{#1}}
\newcommand{\eea}{\end{eqnarray}}

\newcommand{\eeal}[1]{\label{#1}\end{eqnarray}}
\newcommand{\bed}{\begin{deqarr}}
\newcommand{\eed}{\end{deqarr}}
\newcommand{\bedl}[1]{\begin{deqarr}\label{#1}}
\newcommand{\eedl}[2]{\arrlabel{#1}\label{#2}\end{deqarr}}

\newcommand{\beq}{\begin{equation}}
\newcommand{\eeq}{\end{equation}}
\newcommand{\beqa}{\begin{eqnarray}}
\newcommand{\eeqa}{\end{eqnarray}}
\newcommand{\cref}[1]{4\emph{\ref{#1})}}

\newcommand{\RN}{Reissner--Nordstr\"om\ }
\newcommand{\sMP}{standard Majumdar--Papapetrou\ }
\newcommand{\MP}{Majumdar--Papapetrou\ }

\newcommand{\R}{\Bbb R}

\newcommand{\proof}{\noindent{\sc Proof:}\ }

\newcommand{\QED}
   {\hfill$\hbox{\vrule height1.3ex width1.3ex depth.1ex}\ $
     \\ \medskip}
\newcommand{\qed}{\QED}
\newcommand{\eq}[1]{(\ref{#1})}

 \def\scri{\hbox{${\cal J}$\kern -.645em {\raise
      .57ex\hbox{$\scriptscriptstyle (\ $}}}}

\newcommand{\be}{\begin{equation}}
\newcommand{\ee}{\end{equation}}

{\catcode `\@=11 \global\let\AddToReset=\@addtoreset}
\AddToReset{equation}{section}

\newcounter{mnotecount}[section]

\newcommand{\oldmnote}[1]{ \marginpar{\raggedright\tiny\em old mnote
  in the file here  (to be discarded as far as PC is   concerned)}}
\newcommand{\oldnote}[1]{}  
\newcounter{pcheckcount}[section]

\newcommand{\pcheck}[1]{}

\begin{document}

\title{The classification of static  electro--vacuum
  space--times containing an asymptotically flat spacelike
  hypersurface with compact interior}

\author{
Piotr T.\ Chru\'sciel\thanks{Partially supported by a Polish
Research Committee grant 2 P03B 073 24. E-mail
    \protect\url{Piotr.Chrusciel@lmpt.univ-tours.fr}, URL
    \protect\url{ www.phys.univ-tours.fr/}$\sim$\protect\url{piotr}}
  \\ LMPT,
F\'ed\'eration Denis Poisson\\
Tours
  \\
\\
  Paul Tod\thanks{{ E--mail}: paul.tod@st-johns.oxford.ac.uk}
\\
Mathematical Institute and St John's College\\ Oxford}

\maketitle

\begin{abstract}
  We show that static electro--vacuum black hole space--times
  containing an asymptotically flat spacelike hypersurface with
  compact interior and with both degenerate and non--degenerate
  components of the event horizon do not exist. This is done by a
  careful study of the near-horizon geometry of degenerate horizons,
  which allows us to eliminate the last restriction of the
  static electro-vacuum no-hair theory.
\end{abstract}

\section{Introduction}
\label{SI}

A classical question in general relativity, first raised and partially
answered by Israel~\cite{Israel:elvac}, is that of classification of
regular static black hole solutions of the Einstein--Maxwell
equations.  The most complete results existing in the literature so
far are due to Simon~\cite{Simon:elvac},
Masood--ul--Alam~\cite{Masood},
Heusler~\cite{HeuslerRNuniqueness,heuslerMP} and one of us
(PTC)~\cite{Chstaticelvac} (compare Ruback~\cite{Ruback}) leading,
roughly speaking, to the following:
\begin{quote}
\item Suppose that
  \begin{equation}
    \label{eq:qc}
\forall\ i,j\qquad    Q_i Q_j \ge 0\ ,
  \end{equation} where $Q_i$ is the charge of the $i$--th connected
  \emph{degenerate} component of the black hole.  Then the black hole
  is either a Reissner-Norsdstr\"om black hole, or a \MP black hole.
\end{quote}
The above results settled the classification question in the connected
case. The general case, however, remained open.  The aim of this work
is to remove the sign conditions \eq{eq:qc}, finishing the
problem. More precisely, we prove:
\begin{Theorem}
  \label{T1n}
  Let $(M,g,F)$ be a static solution of the Einstein--Maxwell
  equations with defining Killing vector $X$. Suppose that $M$
  contains a connected and simply connected space-like hypersurface
  $\Sigma$, the closure $\bar \Sigma$ of which is the union of an
  asymptotically flat end and of a compact set, such that:
\begin{enumerate}
 \item The Killing vector field $X$ is timelike on $\Sigma$.
\item The topological boundary $\partial \Sigma\equiv
  \overline{\Sigma}\setminus \Sigma$ of $\Sigma$ is a nonempty,
  two-dimensional, topological manifold, with $g_{\mu\nu}X^\mu X^\nu =
  0 $ on $\partial\Sigma$.
  \end{enumerate}
Then, after performing a duality rotation of the electromagnetic field if necessary:
\begin{enumerate}
\item[(i)]  If $\partial \Sigma$ is \emph{connected}, then $\Sigma$ is
  diffeomorphic to $\R^3$ minus a ball.
  Moreover there exists a neighborhood of $\Sigma$ in $M$ which is
  isometrically diffeomorphic to an open subset of the (extreme or
  non--extreme) \RN space--time.
\item[(ii)]  If $\partial \Sigma$ is \emph{not} connected,
  then $\Sigma$ is diffeomorphic to $\R^3$ minus a finite union of
  disjoint balls.  Moreover the space--time contains \emph{only}
  degenerate horizons, and there exists a neighborhood of $\Sigma$ in
  $M$ which is isometrically diffeomorphic to an open subset of the
  \sMP space--time.
\end{enumerate}
\end{Theorem}

The property that the set $\{g_{\mu\nu}X^\mu X^\nu=0\}$ is a
topological manifold,
as well as simple connectedness of $\Sigma$, will hold when
appropriate further global hypotheses on $M$ are made. In fact,
Corollary~1.2 and Theorem~1.3 of \cite{Chstaticelvac}, as well as
the associated remarks, are valid now without the sign
hypothesis~\eq{eq:qc}, and will not be repeated here.

The definitions and conventions used here coincide with those of the
papers~\cite{Chstatic,Chstaticelvac}, except for Section~2 where a
different signature is used.

The idea of the proof is to show that degenerate components of the
horizon are only possible in \sMP space--times, as follows: we start
by showing that the space-metric of a static degenerate horizon is
spherical, with vanishing rotation one-form, and with constant
``second-order surface gravity''. This leads to very precise
information on the geometry of the orbit-space metric near the
horizon. (This part of our work is inspired by the calculations
in~\cite{Reall:2002bh}.) Let $\varphi$ be the electric potential
normalised so that $\varphi$ tends to zero at infinity. One then uses
two conformal transformations of Masood-ul-Alam to prove that this
geometry is possible with $\varphi=\pm 1$ on a component of the
horizon if and only if the metric is a \MP metric. This, together
with~\cite{ChNad} (compare~\cite{HartleHawking}), reduces the problem to one where
$|\varphi|$ is strictly bounded away from one, which has already been
shown to lead to the Reissner-Nordstr\"om geometry
in~\cite{Chstaticelvac}.

\bigskip

\noindent{\sc Acknowledgments:} It is a pleasure to thank M.~Anderson,
R.~Bartnik, H.~Reall and especially H.~Bray for helpful
discussions. The authors acknowledge hospitality and support from the
Newton Institute, Cambridge, during work on this paper.

%
%

\section{The near-horizon geometry of static electrovacuum degenerate Killing horizons}
\label{SPaul} In this section, we establish some results on the
form of the metric of a static electrovac space-time near a degenerate
Killing horizon. There is a range of formalisms available, and we
shall exploit the Newman-Penrose spin-coefficient formalism\footnote{Alternatively, one could introduce the near-horizon
geometry as in~\cite{Reall:2002bh} (compare~\cite{CRT}), and use the discussion of Kundt's
class of metrics in~\cite{ExactSolutions}. We are grateful to H.~Reall
for this observation.} as
reviewed in
\cite{NewmanTod} or \cite{Stewart:book}. To agree with the equations
as they appear in these references, we shall in this section take the
space-time signature to be $(+---)$. As in
\cite{CRT} we introduce Gaussian null coordinates near a component
${\cal{N}}$ of the event horizon, but with the signature changed
so that the metric is
\be
g=r\phi du^2-2dudr-2rh_adx^adu-h_{ab}dx^adx^b.
\label{metric1}
\ee
The Killing vector $X$ is $\partial_u$ with norm
\[g(X,X)=r\phi\]
and ${\cal{N}}$ is located at $r=0$. The surface gravity is
$\kappa=-\partial_{r}(r\phi)$ at $r=0$, and degeneracy of ${\cal{N}}$
means that $\kappa$ vanishes. It follows that
\[\phi=rA(r,x^a)\]
for some $A$. We shall show the following:
\begin{itemize}
\item[(i)]
The co-vector field $h_a$ defined on the spheres $(r=r_0,\;u=u_0)$ vanishes to order $r$.
\item[(ii)]
The metric $h_{ab}$ on the spheres ${\cal{S}}=(r=0,\;u=u_0)$ has
constant Gauss curvature $K$.
\item[(iii)]
On ${\cal{N}}$, $A =K>0$, so that $Ah_{ab}|_{r=0}$ is the unit round
metric on $S^2$.
\item[(iv)] In the purely electric case, the electrostatic potential $\varphi$ satisfies $\partial_r \varphi  = \pm \sqrt{A}$ at $\cal N$.
\end{itemize}
From \cite{CRT} we know that staticity implies that, at ${\cal{N}}$,
$h_a$ is a gradient, say $h_adx^a|_{r=0}=d\lambda$. Now in the metric
(\ref{metric1}), choose the coordinates $x^a$ so that they are
isothermal on ${\cal{N}}$ and then introduce $\zeta=x^1+ix^2$. Choosing  $m$
to be proportional to  $d\bar \zeta$ at $r=0$, the
metric becomes
\be
g=r^2Adu^2-2dudr-2r(hd\zeta+\overline{h}d\overline{\zeta})du-2m\overline{m}\;,
\label{metric2} \ee
where
\begin{eqnarray*}
m&=&-\zX d\bar\zeta+O(r)\;,\\
h&=&\frac{\partial\lambda}{\partial\zeta}+O(r)\;,
\end{eqnarray*}
in terms of functions $\lambda$ (real) and $\zX $ (complex) of
$\zeta$ and $\overline{\zeta}$.

Our goal for the next two pages is equation (2.11). A version of
this equation has appeared in the literature before, as equation
(50) of [11]. In the interest of making the current paper
self-contained, and to introduce notation, we shall rederive it in
the spin-coefficient formalism.
 We will calculate in the null tetrad
$(l^\mu,n^\mu,m^\mu,\overline{m}^\mu)$ by
$$
\begin{array}{lllll}
\l^\mu\partial_\mu&=&D&=&\partial_u +\frac{r^2A}{2}\partial_r\;,\\
n^\mu\partial_\mu&=&\Delta&=&-\partial_r\;,\\
m^\mu\partial_\mu
&=&\delta&=&\frac{1}{\overline{Z}}\partial_{\zeta}+\frac{r}{\overline{Y}}\partial_{\overline{\zeta}}-\left(\frac{rh}{\overline{Z}}+\frac{r^2\overline{h}}{\overline{Y}}\right)\partial_r\;,
\end{array}
$$
where $Z=\zX +O(r)$. Here and elsewhere, a circle over a quantity
indicates the value at $r=0$.

We follow the numbering of \cite{NewmanTod} in the following. The
spin-coefficients are calculated following (A.2) with the
result\footnote{Note that the term $(\alpha-\beta)\pi$ in (A.3g) is misprinted,
and should read $(\alpha-\bar \beta) \pi$.}
\begin{eqnarray*}
\alpha&=&-\frac{1}{2\zX\overline{\zX}}\frac{\partial\overline{\zX}}{\partial\overline{\zeta}}-\frac{1}{4\zX}\overline{h}+O(r)\;,\\
 \beta&=&\frac{1}{2\zX\overline{\zX}}\frac{\partial \zX}{\partial\zeta}-\frac{1}{4\overline{\zX}}{h}+O(r)\;,\\
 \gamma&=&
  -\frac{1}{4}\frac{\partial}{\partial r}\log\left.\left(\frac{Z}{\overline{Z}}\right)\right|_{r=0}+O(r)\;,\\
\epsilon&=&\frac{1}{2}r\zA+O(r^2)\;,\\
\mu&=&-\frac{1}{2}\frac{\partial}{\partial r}\log(Z\overline{Z}) \Big|_{r=0}+O(r)\;,\\
\tau&=&\frac{1}{2\overline{\zX}}h+O(r)\;,
\end{eqnarray*}
together with $\pi = -\overline{\tau},\nu=0,\lambda=1+O(r),\rho=O(r^2), \kappa=O(r^2)$, and $\sigma=O(r^2)$.

From these, we calculate the curvature components from (A.3), setting
the scalar curvature to zero. For the Weyl spinor, we find
$\Psi_0=O(r^2), \Psi_1=O(r), \Psi_3=O(1)$, and $\Psi_4=O(1)$ together
with two expressions for $\Psi_2$:
\begin{eqnarray}
\Psi_2&=&\frac{1}{2\zX\overline{\zX}}\left(\partial_{\zeta}
\partial_{\overline{\zeta}}\lambda-\frac{1}{2}\partial_{\zeta}\lambda\partial_{\overline{\zeta}}\lambda\right)+O(r)\label{psi21}\;,\\
&=&\frac{1}{4}\left(\zA
-K+
\frac{1}{2\zX\overline{\zX}}\partial_{\zeta}\lambda\partial_{\overline{\zeta}}\lambda\right)+O(r)\label{psi22}
\end{eqnarray}
where $K=-\frac{1}{\zX \overline{\zX
}}\partial_{\zeta}\partial_{\overline{\zeta}}\left(\log(\zX
\overline{\zX})\right)$, which is the Gauss curvature of
${\cal{S}}$.

For the Ricci spinor, we find $\Phi_{00}=O(r^2), \Phi_{01}=O(r)$ and
the remaining components are $O(1)$. In particular we have
\begin{eqnarray}
\Phi_{11}&=&\frac{1}{4}\left(\zA +K+
\frac{1}{2\zX\overline{\zX}}\partial_{\zeta}\lambda\partial_{\overline{\zeta}}\lambda\right)+O(r)\label{phi11}\;,\\
\Phi_{02}&=&\frac{1}{2\overline{\zX}^2}\left(\partial_{\zeta}\partial_{\zeta}\lambda-\frac{\partial_{\zeta}(\zX\overline{\zX})}{(\zX\overline{\zX})}\partial_{\zeta}\lambda-\frac{1}{2}(\partial_{\zeta}\lambda)^2\right)+O(r)\;.\label{phi02}
\end{eqnarray}
Since we are concerned with electrovac solutions, the Ricci spinor
$\Phi_{ABA'B'}$ is obtained from the Maxwell spinor $\phi_{AB}$
according to
\[\Phi_{ABA'B'}=k\phi_{AB}\overline{\phi}_{A'B'}\;,\]
where $k=\frac{2 G}{c^4}$. (We shall often  assume $G=c=1$.)
 In particular this means that
\be
\Phi_{00}=k\phi_0\overline{\phi}_0;\;\Phi_{02}=k\phi_0\overline{\phi}_2;\;\Phi_{11}=k\phi_1\overline{\phi}_1\;.\label{phis}
\ee
We saw above that $\Phi_{00}=O(r^2)$ 
so
that, from the first equation in (\ref{phis}), we deduce that $\phi_0=O(r)$ and
since $\phi_2=O(1)$ we must have $\Phi_{02} = O(r)$. By (\ref{phi02})
this is
\be
\partial_{\zeta}\partial_{\zeta}\lambda-\frac{\partial_{\zeta}(\zX\overline{\zX})}{(\zX\overline{\zX})}\partial_{\zeta}
\lambda-\frac{1}{2}(\partial_{\zeta}\lambda)^2=0.
\label{lam1}
\ee
A second equation on $\lambda$ follows from (\ref{psi21}), (\ref{psi22}) and (\ref{phi11}) as
\be
\frac{1}{2\zX\overline{\zX}}\left(\partial_{\zeta}
\partial_{\overline{\zeta}}\lambda-\frac{1}{2}
\partial_{\zeta}\lambda\partial_{\overline{\zeta}}\lambda\right)
=k\phi_1\overline{\phi}_1-\frac{1}{2}K.
\label{lam2}
\ee
The component $\phi_1$ of the Maxwell field is constrained by the
Maxwell equations, specifically by (A.5b) of \cite{NewmanTod} which
here becomes
\[\overline{\delta}\phi_1+2\pi\phi_1=O(r)\;,\]
or
\be
\partial_{\overline{\zeta}}\phi_1
-\phi_1\partial_{\overline{\zeta}}\lambda=O(r)\;.\label{phi3}
\ee
This integrates at once to give $\phi_1=\zphione e^{\lambda}+O(r)$
where $\zphione $ is holomorphic in $\zeta$ on ${\cal{S}}$. It is
also bounded (since it is the contraction of the self-dual part of
the Maxwell field with the volume form of ${\cal{S}}$), and so it
must be constant (the value of this constant is proportional to
the charge of the black hole). We use this in (\ref{lam2}), and
then we can write (\ref{lam1}) and (\ref{lam2}) jointly as a
tensor equation on ${\cal{S}}$ as
\be
\nabla_a\nabla_b\lambda-\frac{1}{2}\nabla_a\lambda\nabla_b\lambda=-\zR
_{ab}+2k|\zphione |^2e^{2\lambda}\zh _{ab} \label{lam3} \ee
where, as before, a circle over a quantity indicates the value at
$r=0$. As noted above, this is equivalent to equation (50) of
\cite{LP1} (also our (\ref{phi3}) is equivalent to their (47)). We
shall deduce from (\ref{lam3}) that necessarily $\lambda$ is
constant and $\zh _{ab}$ is the metric of a round sphere.

First, introduce $\psi=e^{-\lambda/2}$ so that (\ref{lam3}) becomes
\be
\nabla_a\nabla_b\psi=\frac{\psi}{2}\zR _{ab}-k
|\zphione|^2\psi^{-3}\zh _{ab} \label{psi1}\;, \ee and from the trace of
this find
\be
\Delta\psi=\frac{\psi}{2}\zR -2k|\zphione |^2\psi^{-3}\;.
\label{psi2} \ee
Take $\nabla^b$ on (\ref{psi1}) and use (\ref{psi2}) to find
\[\nabla_a\left(\psi^3\zR -12k|\zphione |^2\psi^{-1}\right)=0\;,
\]
so that
\be
\zR =\frac{c_1}{\psi^3}+\frac{12k}{\psi^4}|\zphione |^2 \label{ric}
\ee for some constant $c_1$. Insert this into \eq{psi2}, and then
into \eq{psi1}, to obtain
\begin{eqnarray}\label{psi2.1}
&\displaystyle \Delta\psi=\frac{c_1\psi+8k|\zphione |^2}{2\psi^3}\;,
&\\&\displaystyle\nabla_a\nabla_b\psi=
\frac{c_1\psi+8k|\zphione |^2}{4\psi^3}\zh_{ab}\;.&\label{psi2.2}
\end{eqnarray}
The possibility that $c_1\ge 0$ or $\zphione =0$ leads to $\Delta
\psi$ of constant sign, which is possible on a compact manifold only
if $c_1=\zphione =0$ and $\psi$ is constant, leading to
$\zR_{ab}=0$. But there are no such metrics on $S^2$, hence \bel{cone}
c_1<0 \ \mbox{ and } \ \zphione\ne 0\;.\ee

Multiply (\ref{psi1}) by $\nabla_b\psi$ and integrate using
(\ref{ric}) to find
\be
|\nabla\psi|^2=c_2-\frac{c_1}{2\psi}-\frac{2k}{\psi^2}|\zphione |^2
\label{norm} \ee for some constant $c_2$.

In order to analyse the critical set of $\psi$, consider any geodesic
$\gamma(s)$, $s\in I$,  with unit tangent $\dot \gamma$. From
\eq{psi2.2} we find $$\frac {d^2 \psi}{ds^2}=
\frac{c_1\psi+8k|\zphione |^2}{4\psi^3}\;.$$ Suppose that $p$ is a
critical point of $\psi$ and  that the right-hand-side vanishes at
$p$.
Then $\psi(\gamma(s))=\psi(p)$ is a solution satisfying the right
initial data at $p$, and uniqueness of solutions of ODEs shows then
that $\psi$ is constant on all geodesics through $p$. It easily
follows from \eq{psi1} and \eq{psi2.1} that both $\psi$ and the metric
are analytic in an appropriate chart, so this situation arises if and
only if $\psi$ is constant. Supposing it is not, we conclude that the
Hessian of $\psi$ is strictly definite at critical points, and the
Laplacian does not vanish there. Consequently, the critical set of
$\psi$ is either a finite collection of points, or the whole
sphere. In the former case, each critical point is a strict local
extremum. In both cases the set \bel{Omsetdef} \Omega:=\{\nabla \psi\ne 0\}\ee is connected (possibly empty).

We next show that $c_2$ is negative. Let $$a:= \inf \psi\;, \quad b:=\sup \psi\;,$$ then $0<a\le
b<\infty$ since $\psi$ is positive and smooth. Suppose that $a\ne
b$. At any $p$ such that $\psi(p)=b$ we have $\nabla \psi (p)=0$
and $\Delta \psi (p)< 0$.\footnote{The inequality has to be
strict, otherwise $\psi$ is constant, either by the geodesic
argument just given, or by the fact that $\Delta \psi$ would again
have constant sign.} The latter together with \eq{psi2.1} gives
$c_1b < -8k|\zphione |^2$, while the former reads, in view of
\eq{norm},
$$c_1 b = 2c_2b^2 - 4k|\zphione |^2\;,$$ leading to $$c_2b^2 < -2k
|\zphione |^2\;,$$ implying $c_2< 0$.

Our next target is to derive \eq{locform} below.
Let $p_-$ be a minimum of $\psi$ and let $\gamma$ be any geodesic
starting at $p_-$, with tangent $\dot \gamma$ of unit length. Then
$\psi\circ \gamma$ is a solution of the Cauchy problem
$$\frac {d^2 \psi}{ds^2}=
\frac{c_1\psi+8k|\zphione |^2}{4\psi^3}\;,\quad \psi(0)=a\;,\quad
\frac {d\psi}{ds}(0)=0\;,$$
 which shows that $\psi$ depends only upon the geodesic distance from $p_-$, and not on the
 direction of the geodesic. Thus,
 the level sets of $\psi$ coincide with the geodesic spheres
 centred at $p_-$, within the injectivity radius of $p_-$.
 A similar conclusion holds at any maximum of $\psi$.

 On
$\Omega$, as defined in \eq{Omsetdef}, we may use $\psi$ locally as a coordinate, leading to the
following form of the metric
\[\frac{d\psi^2}{F^2(\psi)}+H^2(\psi,\phi)d\phi^2\;,\]
where $\phi$ is a local coordinate on the level sets of $\psi$,
and
\be
F^2(\psi)=|\nabla
\psi|^2=c_2-\frac{c_1}{2\psi}-\frac{2k}{\psi^2}|\zphione |^2.
\label{F} \ee
Equation (\ref{psi2}) implies $H^2=F^2(\psi)G(\phi)$ and we may
redefine $\phi$ to make $G=1$, leading to the local form
\bel{locform} \frac{d\psi^2}{F^2(\psi)}+F^2(\psi)d\phi^2\ee
\newcommand{\cthree}{c_3}%
Within the radius of injectivity of $p_-$ and away from $p_-$ we
have $d\psi/F=d\rho$, where $\rho$ is the distance function from
$p_-$. Note that
\bel{rho1} F^2(\psi)=\cthree \rho^2+o(\rho^2)
 \ee
for small $\rho$, for a constant $c_3$ which can be read off from
\eq{F}; this is compatible with elementary regularity provided that
$\phi$ has appropriate periodicity.

The integral curves of $\nabla \psi$ can be used to obtain a
diffeomorphism between the level sets of $\psi$ within $\Omega$, which
shows that
\eq{locform} provides a global representation of the metric on $\Omega$.

Let $p_+$ be any point in $\cal S$ such that $\psi(p_+)=b$, then $p_+\in
\overline \Omega$, and likewise the level sets of $\psi$ near $p_+$
are geodesic spheres. This, and what has been said, implies that the
set $${\hat {\cal S}}:=\{p_-\}\cup \Omega\cup \{p_+\}$$ is both open
and closed in $\cal S$, hence $\hat {\cal S} =\cal S$. Furthermore,
 within the radius of injectivity of $p_+$ and away from $p_+$, we have
$d\psi/F=d\hat \rho$, where $\hat \rho$ is the distance function
from $p_+$. Since the periodicity of $\phi$ has already been
determined we must have
 \bel{rho2} F^2(\psi)=\cthree^2\hat \rho^2+o(\hat
\rho^2)
 \ee
for small $\hat \rho$. Eliminating $\cthree$ between \eq{rho1} and
\eq{rho2}, a standard calculation leads to
  $$(F^2)'(a)=-(F^2)'(b)\;.$$
  Equivalently,
  $$c_1ab(a^2+b^2)=-
{8k}|\zphione |^2(b^3+a^3)\;.$$ Eliminating $c_2$ from the
equations $F(a)=F(b)=0$ one finds
$$c_1ab=-
{4k}|\zphione |^2(a+b)\;.$$ Substitute
into the previous equation
to obtain $a=b$, which is a contradiction.

 We  conclude that regularity of the
metric of ${\cal{S}}$ requires that $|\nabla\psi|=0$, so $\psi$
and therefore $\lambda$ are constant, which establishes $(i)$.
From (\ref{lam3}), if $\lambda$ is constant then the metric $\zh
_{ab}$ is that of a round sphere, establishing $(ii)$. Next,
from (\ref{psi21}), with $\lambda$ constant, $\Psi_2$ is zero at
${\cal{N}}$ and then from (\ref{psi22}), $\zA =K$, establishing
$(iii)$.

Finally, recall that the electrostatic potential $\varphi$ is defined
by the equation $$d\varphi = i_X F\;,$$ where $F$ is the Maxwell
two-form and $i_X$ denotes contraction with $X$.  Since $X= l +r^2
An/2$, in the purely electric case we have $$\varphi_1 = \frac 12
F_{ab}l^a n^b = \frac 12 \frac{\partial \varphi}{\partial r}\;.$$ But
(see \eq{phis} and \eq{phi11}) $$\Phi_{11} = 2 \varphi_1 \bar\varphi_1
=\frac 12 \zA +O(r)\;,$$ so $\partial_r
\varphi = \pm \sqrt{A}$ at $\cal N$, establishing $(iv)$.


\section{Proof of Theorem \protect\ref{T1n}}
\label{ptt1n}

As argued by Heusler~\cite{HeuslerRNuniqueness}
(compare~\cite[Lemma~3.2]{Chstaticelvac}), a duality rotation
guarantees that the Maxwell field is purely electric.
Following~\cite{Chstatic}, we equip $\Sigma$ with the orbit space
metric\footnote{In~\cite{Chstatic} the symbol $h$ is used; to avoid a
clash of notation with the previous section we  use $\gamma$
instead.} $\gamma$ defined as
\begin{equation}
\gamma(Y,Z)=g(Y,Z)- \frac{g(X,Y)g(X,Z)}{g(X,X)}\ ,
 \label{eq:hdefnew}
\end{equation}
where $X$ is the defining Killing vector, that is, the Killing
vector which asymptotes $\partial/\partial t$ in the asymptotic
regions, and satisfies the staticity condition. 

As in~\cite{Masood}, we consider the functions
\bel{Omdef}
\Omega_\pm =\frac {(1\pm V)^2-\varphi^2}4\;,
 \ee
where $-V^2$ is the Lorentzian length of the Killing vector $X$,
 and the metrics
 \bel{gpmdef}g_\pm:=\Omega_\pm^2 \gamma \;.\ee
(The interest in those metrics arises from the positivity of their
scalar curvatures~\cite{Masood}.) Proposition~3.4
of~\cite{Chstaticelvac} shows that
 \bel{maxprinineq} 0\le
|\varphi| \le 1-V\;,\ee hence the functions $\Omega_\pm$ are
non-negative, with the inequalities being strict in the interior
unless the metric is locally a \MP metric (compare~\cite{Masood}).
From now on we suppose that this is \emph{not} the case. The
possibility that $|\varphi| $ is strictly bounded away from one
leads to the Reissner-Nordstr\"om solutions~\cite{Chstaticelvac},
so we assume, for contradiction, that there exists a component of
the horizon ${\zmcH}$ on which
$\varphi|_{r=0}=:\epsilon\in\{\pm1\}$, then ${\zmcH}$ is
degenerate~\cite[Prop.~3.4]{Chstaticelvac}. By the results in
Section~\ref{SPaul} and by \eq{maxprinineq} we have
 $$\varphi=\epsilon(1-\sqrt{\zA }r+O(r^2))\;,$$ and since
$V=\sqrt{|g_{uu}|}=\sqrt{\zA }r+O(r^2)$ we obtain
\beal{Omdev1} \Omega_+&=& \frac 14\Big( (1+\sqrt{\zA }r+O(r^2))^2-
(1-\sqrt{\zA }r+O(r^2))^2\Big)=\sqrt{\zA
}r+O(r^2)\;,\phantom{xxxx}
\\ \Omega_-&=& \frac 14\Big(
(1-\sqrt{\zA }r+O(r^2))^2-(1-\sqrt{\zA }r+O(r^2))^2\Big)=O(r^2)\;.
 \eeal{Omdev2}

We can use the space-times coordinates $r$ and $x^a$ of
Section~\ref{SPaul} as coordinates on the orbit space near
$\{r=0\}$. From \eq{eq:hdefnew} and Section~\ref{SPaul} we infer
\beal{eq1} \gamma_{rr}&=& g_{rr} - \frac{g_{ru}^2}{g_{uu}}=
\frac{1}{\zA r^2+O(r^3)}\;,
\\
\gamma_{ra}&=& g_{ra} - \frac{g_{ru}g_{au}}{g_{uu}}= \frac{O(r^2)}{\zA
r^2+O(r^3)}\;, \label{eq2}
 \\
\gamma_{ab}&=& g_{ab} - \frac{g_{au}g_{bu}}{g_{uu}}=\zh_{ab}+O(r)+
\frac{O(r^4)}{\zA r^2+O(r^3)}\;.\eeal{eq3} This leads to the
following form of the metric $g_+$:
 \bean g_+&=& \frac{\Omega_+^2}{g_{uu}} \times g_{uu} \gamma  =\frac{\Omega_+^2}{g_{uu}}
  \left(dr^2+O(r^2)dx^a dr+ (g_{uu}g_{ab}+O(r^4))
dx^adx^b\right)
 \\
 &=& \Big(1+O(r)\Big)\left(dr^2+O(r^2)dx^a dr+ \Big(r^2(1+O(r))\zA\zh_{ab}+O(r^3)\Big)
dx^adx^b\right)\;.
 \nonumber
 \\ && \eeal{eq2a}
    We want to think of the coordinate $r$ above as a radial
    coordinate near the origin of $\R^3$. First, since $\zA\zh_{ab}$
    is the unit round metric on $S^2$ we have
    $$r^2\zA\zh_{ab}dx^adx^b= \sum (dx^i)^2 -dr^2\;,$$ so this part of
    the metric combines with the leading part of the $dr^2$ term in
    \eq{eq2a} to give a smooth tensor field.  Next, the form $rdr=\sum
    x^i dx^i$ is smooth with respect to the standard differentiable
    structure on $\R^3$, and vanishes at the origin as $O(|\vec x|)$,
    so that the term $O(r)dr^2$ gives a contribution which, in the
    coordinates $x^i$, vanishes at the origin as $O(|\vec x|)$, with
    bounded first derivatives, and second derivatives dominated by a
    multiple of $r^{-1}$.  To understand the remaining terms, a
    seemingly straightforward approach is to use spherical coordinates
    on $S^2$. However, those coordinates are singular at the $z$-axis,
    which leads to problems when one wishes to capture the regularity
    of the resulting metric. An alternative way of handling this
    proceeds as follows:

Think of the sphere $S^2$ as a subset of $\R^3$ with global
coordinates $\hat x^i$, and let $\beta$ be any smooth one-form on
$S^2$. Then $\beta$ can be uniquely extended to a smooth one-form
$\hat \beta_i (\hat x^j) d\hat x^i$ defined on $\R^3\setminus
\{0\}$ by requiring that
$$\hat x^i \partial_{\hat x^i} \hat \beta_j =0\;,\quad \hat \beta_j
\hat x^j =0\;,\quad i^*_{S^2}(\hat \beta_i d\hat x^i)=\beta\;,$$
where $i^*_{S^2}$ is the pull-back map. Similarly any
two-covariant tensor field $\alpha$ on $S^2$ can be uniquely
extended to a smooth tensor field  $\hat \alpha_{k\ell} (\hat x^j)
d\hat x^k d\hat x^\ell $ defined on $\R^3\setminus \{0\}$ by
requiring that
$$\hat x^i \partial_{\hat x^i} \hat \alpha_{k\ell} =0\;,\quad \hat \alpha_{k\ell}
\hat x^k = \hat \alpha_{k\ell} \hat x^\ell =0\;,\quad
i^*_{S^2}(\hat \alpha_{k\ell} d\hat x^k d\hat x^\ell )=\alpha\;.$$

 Let $B^*(\vec 0,a):=B(\vec 0,a)\setminus\{\vec 0\}$ denote a
  punctured coordinate ball centred at $\vec 0$, of
 radius $a$, in
$\R^3$ and consider the map \begin{eqnarray*} \Psi: B^*(\vec 0,a)
&\to& (0,\infty)\times \R^3\;,
 \\
 x^i &\mapsto & \left(r=\sqrt{\sum (x^i)^2}, \hat x^i=\frac{x^i} r\right)\;.
 \end{eqnarray*}
A term $\beta_a  dx^a dr$ in the metric extends as
above to a tensor field $ \hat \beta_i d\hat x^i dr$ on
$(0,\infty)\times \R^3$, and its pull-back by $\Psi$ produces a
term
$$\Psi^*( \hat \beta_i d\hat x^i) = \hat \beta_i d\left(\frac
{x^i} r\right) =\hat \beta_i \frac {dx^i} r- \underbrace{\hat \beta_i
\frac {x^i} {r^2}}_{=0} dr=\hat \beta_i \frac {dx^i} r\;.$$ This
shows that the terms $O(r^2)dx^a dr$ in \eq{eq2a} give
contributions, in the coordinates $x^i$, of the form
$$r \times \Big(\mbox{smooth function of }\ \frac {x^i} r \Big) dx^j dx^k\;.$$
 A similar analysis of the remaining $dx^a dx^b$ terms shows that, in the coordinates $x^i$, the
 metric $g_+$ can be extended by continuity through the origin to a metric
still denoted by the same symbol, of the form
 \bel{uniel}g_+=(\delta_{ij}+O(|\vec x|))dx^i dx^j\;,\ee
 with derivatives satisfying, for some constant $C$,
 \bel{Coo} |\partial_j (g_+)_{k\ell}|\le
 C\;,\quad |\partial_i\partial_j (g_+)_{k\ell}|\le
 C|\vec x|^{-1} \;.\ee



The key fact in the remainder of the proof is  the following: the
positivity of the scalar curvatures of both metrics $g_{\pm}$
implies a differential inequality on the the quotient
$\Omega_-/\Omega_+$, which is incompatible with the vanishing of
this quotient at the origin. Some technicalities are required
because of the potential lack of smoothness of $g_\pm$ at the
origin. We start by rescaling $g_+$ so that the scalar curvature
vanishes:

\begin{Proposition}
\label{Pconffa} There exists $b>0$ and a
 positive function   $\psi\in C(B(\vec 0,b))\cap C^\infty(B^*(\vec
 0,b))$,
 bounded away from zero, such that
 the scalar curvature of $\psi^4 g_+$ vanishes on $B^*(\vec
 0,b)$.
\end{Proposition}

\proof  We want to construct a solution $\psi$ in
$B^*(\vec 0,b)$  of the  equation
 $$R(\psi^4 g_+)\psi^5=-8\Delta_{g_+} \psi +  {R(g_+)} \psi =
 0\;.$$ We look for $\psi$ of the form $\psi=1+u$, where $u$ vanishes on
a coordinate sphere of radius $b$. The equation for $u$ reads
 \bel{ueq} -8\Delta_{g_+} u +  {R(g_+)} u =-{R(g_+)}\;.\ee
From \eq{uniel}-\eq{Coo} one finds
that the scalar curvature $R(g_+)$ of the metric $g_+$ satisfies
\bel{Coo2} R(g_+)\le \frac{C'}{|\vec x|}\ee
for some constant $C'$. By scaling $\vec x \to b^{-1} \vec x$ we can
assume $b=1$, note that \eq{Coo2} becomes then \bel{Coo3} R(g_+)\le
\frac{C'b}{|\vec x|}\;.\ee It follows that the right-hand-side of
\eq{ueq} is in $L^2(B(\vec 0,1))$. To solve \eq{ueq}
one can proceed as follows: Let $0<\epsilon<1$; we wish, first, to
show the existence of a solution $u_\epsilon\in C^\infty(B(\vec
0,1)\setminus B(\vec 0,\epsilon) $ of \eq{ueq} that vanishes both on
the coordinate sphere of radius one and on that of radius
$\epsilon$. This will follow from the standard theory if we can show
that the solutions of the homogeneous equation, still denoted by
$u_\epsilon$ are unique.  For this, extend $u_\epsilon$ by zero to the
interior ball of radius $\epsilon$, and recall the Hardy inequality
$$\int_ {B(\vec 0,1)} \frac {u^2}{r^2} \le C \int_ {B(\vec 0,1)}
|du|^2 $$ (note that the standard version thereof uses the flat
metric, but by uniform ellipticity both the measure and the norm of
$du$ can be taken with respect to the current metric with an appropriately modified  constant $C$). We then have
$$\int_ {B(\vec 0,1)} {|R(g_+)|} {u^2} \le C'b \int_ {B(\vec 0,1)}
\frac {u^2}{r} \le CC'b\int_ {B(\vec 0,1)} |du|^2 \;.$$ We can choose
$b$ small enough to obtain $CC'b\le 1$, then \begin{eqnarray*}
0&=&\int_ {B(\vec 0,1)\setminus B(\vec 0,
\epsilon)}u_\epsilon(-8\Delta_{g_+} u_\epsilon + {R(g_+)} u_\epsilon)
\\ &=&\int_
{B(\vec 0,1)\setminus B(\vec 0, \epsilon)}8|du_\epsilon|^2 + {R(g_+)} u_\epsilon^2
\ge 7\int_ {B(\vec 0,1)\setminus
B(\vec 0, \epsilon)}|du_\epsilon|^2\;,\end{eqnarray*} giving uniqueness, as
desired.

In the case of the non-homogeneous equation the last calculation
further gives
\begin{eqnarray*} 7\int_ {B(\vec 0,1)}|du_\epsilon|^2&\le&
\int_
{B(\vec 0,1)}u_\epsilon(-8\Delta_{g_+}
u_\epsilon + {R(g_+)} u_\epsilon)=-\int_
{B(\vec 0,1)} {R(g_+)} u_\epsilon
\\ & \le &  \frac 1 {2\epsilon} \int_
{B(\vec 0,1)} {(R(g_+)r)^2} + \frac  {\epsilon}2 \int_
{B(\vec 0,1)} \frac{u_\epsilon^2}{r^2}
\\ & \le &  \frac 1 {2\epsilon} \int_
{B(\vec 0,1)} {(R(g_+)r)^2} + \frac  {C\epsilon}2 \int_
{B(\vec 0,1)} |d{u_\epsilon^2}|
 \;,
\end{eqnarray*}
where we have used $2xy \le \epsilon^{-1} x^2+\epsilon y^2$. Choosing
$\epsilon= 2C^{-1}$, we can carry the last term to the left-hand-side,
which shows that the sequence $u_\epsilon$ is bounded in
$H^1$. Standard arguments imply the existence of a function $u$
solving \eq{ueq} on $B^*(\vec 0,1)$. We can use Sobolev's inequality
and \cite[Corollary~1.1, p.~29]{Veronbook} to conclude that $u$ is a
weak solution of
\eq{ueq} on $B(\vec 0,1)$. By the last calculation one has $\Delta_{g_+}
u\in L^2$ (in a weak sense), and since the metric $g_+$ is Lipschitz continuous we can invoke
elliptic theory \cite[Theorem~8.8]{GilbargTrudinger77} to conclude that $u\in
H^2\subset C^0$. Since all the norms involved can be made arbitrarily
small when $b$ is small enough, we will have $\|u\|_{L^\infty}\le 1/2$
for appropriate $b$, hence $\psi\ge 1/2$. \qed

 We turn our attention now to the metric $g_-$; as already pointed out,
the sign of its scalar curvature will provide the desired
contradiction.  In the coordinates
 $\vec x$ introduced above, on $B^*(\vec 0,b)$ we have $$g_-=
 \Omega_-^2 \gamma = {\left(\frac {\Omega_-^2 }{\Omega _+^2}\right)}
 g_+= \underbrace{\left(\frac {\Omega_-^2 }{\Omega
 _+^2}{\psi^{-4}}\right)}_{=:\hat \psi^4}\underbrace{{\psi^4}
 g_+}_{=:\hat g} \;.  $$ with $\hat \psi$ nonnegative, and vanishing
 precisely at the origin. The transformation law of the scalar
 curvature under conformal transformation gives, on $B^*(\vec 0,b)$,
 $$-8\Delta_{\hat g} \hat \psi +\underbrace{ {R(\hat g)}}_{=0} \psi =
 R(g_-)
\hat \psi^5\ge 0\;.
 $$

  Let $H$ be the $\Delta_{\hat g}$--harmonic function in $B(\vec
  0,b)$ which equals $\hat \psi$ on the boundary of the ball. The function $H$ can be constructed by minimising $$\int_{B(\vec 0,b)}|dH|^2_{\hat g}d\mu_{\hat g}$$ in the class of functions satisfying the boundary data, hence $H\in H^1(B(\vec 0,b))$. By the
  maximum principle~\cite[Theorem~8.1]{GilbargTrudinger77} (note that
  uniform ellipticity of the metric, guaranteed by
\eq{uniel}, and the $H^1$ character of $H$, suffice for this)  $H$ is bounded away
from zero.

Applying the maximum principle on the set $B(\vec 0,b) - B(\vec 0,e)$
for $e < b$ tells us that $\hat \psi$ is greater than or equal to the
harmonic function which equals $\hat \psi$ on the outer boundary
$S_b(\vec 0)$ and zero on the inner boundary $S_e(\vec
0)$.\footnote{The comparison argument in this paragraph has been
pointed out to us by H.~Bray.}  In the limit as $e$ goes to zero, this
harmonic function converges, uniformly on compact subsets, to a
bounded function $\hat H$ which solves the Laplace equation on
$B^*(\vec 0,b)$. By Serrin's removable singularity
theorem~\cite[Theorem~1.19, p.~30]{Veronbook} (note again that
\eq{uniel}, suffices for this) it holds that $\hat
H=H|_{B^*(\vec 0,b)}$.  Thus, $\hat \psi \ge H(x)$ in $B^*(\vec
0,b)$. Since $H(x)$ is strictly positive in $B(\vec 0,b)$, and $\hat
\psi(z)$ is continuous at $\vec 0$, by the comparison
principle~\cite[Theorem~8.1]{GilbargTrudinger77} we obtain $\hat \psi(\vec 0) > 0$.
This contradicts the fact that $\Omega_-/\Omega_+$ tends to zero as
$\vec x$  approaches the origin. Hence $\varphi(\vec 0)=\pm 1$ is only possible for \MP solutions, and the remarks at the end of the Introduction
complete the proof.

 \bibliographystyle{amsplain}
%
\bibliography{../references/newbiblio,%
../references/reffile,%
../references/bibl,%
../references/hip_bib,%
../references/newbib,%
../references/PDE,%
../references/netbiblio,%
BCT}

%
%



\end{document}